\documentclass[twocolumn,
amssymb, amsmath, prd]{revtex4}
\usepackage{graphicx}

\begin{document}

\title{SAXSFit: A program for fitting small-angle x-ray and neutron scattering data}

\author{Bridget Ingham}
\email[Corresponding author: ]{b.ingham@irl.cri.nz}
\affiliation{Industrial Research Limited, P. O. Box 31-310, Lower
Hutt 5040, New Zealand}

\author{Haiyong Li}
\affiliation{Department of Chemical and Materials Engineering, San
Jose State University, San Jose, CA 95192, USA}

\author{Emily L. Allen}
\affiliation{Department of Chemical and Materials Engineering, San
Jose State University, San Jose, CA 95192, USA}

\author{Michael F. Toney}
\affiliation{Stanford Synchrotron Radiation Lightsource, 2575 Sand
Hill Road, Menlo Park, CA 94025, USA}

\date{October 13, 2008}

\begin{abstract}

SAXSFit is a computer analysis program that has been developed to
assist in the fitting of small-angle x-ray and neutron scattering
spectra primarily from nanoparticles (nanopores). The fitting
procedure yields the pore or particle size distribution and eta
parameter for one or two size distributions (which can be
log-normal, Schulz, or Gaussian). A power-law and/or constant
background can also be included. The program is written in Java so
as to be stand-alone and platform-independent, and is designed to be
easy for novices to use, with a user-friendly graphical interface.

\end{abstract}


\maketitle

\section{Introduction}

Small-angle x-ray scattering (SAXS) and small-angle neutron
scattering (SANS) are well-established and widely used techniques
for studying inhomogeneities on length scales from near-atomic scale
(1 nm) up to microns (1000 nm). Recently, there has been an
increasing emphasis and importance of nanoscale materials, due to
the distinct physical and chemical properties inherent in these
materials \cite{Pedersen2002,Frazel2003}. This, together with the
significant advances in X-ray and neutron sources, has resulted in
the dramatically increased use of SAXS and SANS for characterizing
nanoscale materials and self-assembled systems. For example, these
techniques are used to investigate polymer blends, microemulsions,
geological materials, bones, cements, ceramics and nanoparticles.
These measurements are often made over a range of length scales and
in real time during materials processing or other reactions such as
synthesis. However, there has been less progress in SAXS and SANS
data analysis, although some analysis software is available. For
example, programs based on IGOR Pro primarily for the reduction and
analysis of SANS and ultra-small-angle neutron scattering (USANS)
are available from NIST \cite{Kline2006}. PRINSAS has been developed
for the analysis of SANS, USANS and SAXS data for geological samples
and other porous media \cite{Hinde2004}. PRIMUS and ATSAS 2.1 are
used primarily for the analysis of biological macromolecules in
solutions, but can be used for other systems such as nanoparticles
and polymers \cite{Konarev2003, Konarev2006}. FISH is another SANS
and USANS fitting program developed at ISIS \cite{Heenan2007}. The
Indra and Irena USAXS data reduction and analysis package developed
at APS \cite{Ilavsky2008} is also based on IGOR. Both of these
latter programs offer several advanced features, including multiple
form factor choices and background reduction routines. While these
programs provide a powerful analysis capability, they can be
complicated to use and some are based on commercial software. This
has motivated the development of a simple, easy to use analysis
package.

In this paper, we describe SAXSFit - a program developed to fit SAXS
and SANS data for systems of particles or pores with a distribution
of particle (pore) sizes. SAXSFit is easy to use and applicable to a
wide variety of materials systems. The program is most appropriate
to low concentrations of particles or pores, due to the
approximations used, but it does account for interparticle
scattering within the local monodisperse approximation
\cite{Pedersen1994}. The program is based on Java and is readily
portable with a user-friendly graphical interface. The emphasis of
SAXSFit is to provide an easy-to-use analysis package primarily for
novices, but also of use to experts.

\section{Software description and use}

\begin{figure*}
\includegraphics*[width=170mm]{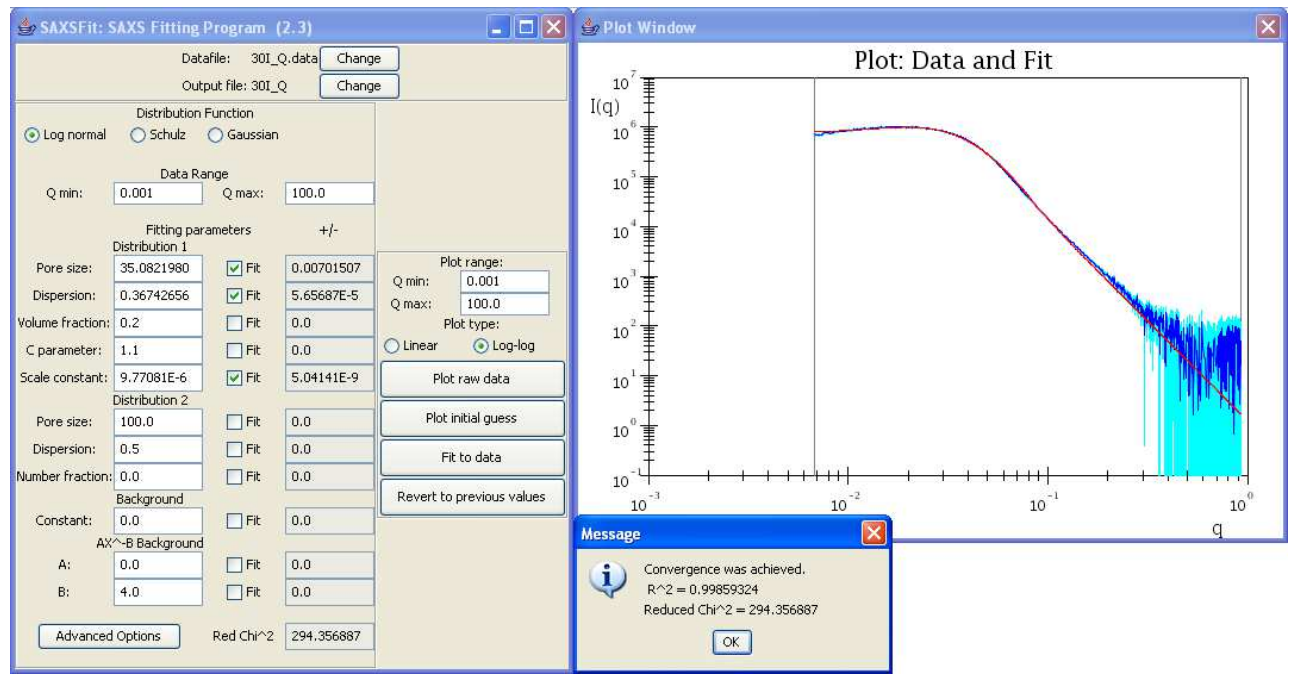}
\caption{\label{fig1} Screen dump of the SAXSFit graphical user
interface and plot window.}
\end{figure*}

SAXSFit is written in Java (SDK 1.4.2) and provides a graphical user
interface (Figure \ref{fig1}) to select and adjust parameters to be
used in the fit, change the plotting display and range of data to be
used, calculating `initial guess' patterns and running the fit. It
uses the algorithms of a Matlab-based program \cite{Li2004}. The
advantage of using Java is to provide a stand-alone program which is
platform-independent, along with having a user-friendly graphical
interface. SAXSFit is also available as a Windows executable.

SAXSFit can read ASCII data files (comma, space, or tab delimited),
with or without a non-numerical header, which consist of two column
($q$, $I(q)$) or three-column ($q$, $I(q)$, error bars) data. Any
subsequent columns in the data file are ignored. Once a data curve
is successfully imported from an input file, it is plotted in a
separate window and the fitting buttons are enabled. The error bars
are also plotted if the input file contains them. The plot can be
manipulated by changing the $q$-range and selecting whether it is
log-log or linear. Initial guess and fitted curves are displayed
when they are calculated. The $q$-min and $q$-max values of the data
to be fitted are also shown as vertical lines and can be altered by
changing the appropriate text boxes. Several fitting parameters are
available, with the option to fit or fix their values.

Three distributions of particle/pore sizes are available:
log-normal, Schulz and Gaussian. These use the same two fitting
parameters, `particle/pore size' ($r_0$) and `dispersion'
($\sigma$), and are detailed in Section \ref{sec3-1}. The units for
the pore size are the inverse of the units of the data (i.e. {\AA}
for data in {\AA}$^{-1}$, or nm for data in nm$^{-1}$). A second
size distribution can also be included in the fit. It has been shown
that the choice of distribution function does not dramatically
affect the final result \cite{Caponetti1993, Kucerka2004}. A
constant background and/or power law $Aq^{-B}$ can also be included.

Advanced options include the ability to change the maximum number of
iterations and the weighting of the data. Again there are several
options: a constant weighting ($w = 1$), statistical weighting ($w =
1/I(q)$), or uncertainty weighting ($w = 1/\Delta I(q)^2$ - only
applicable where data error bars ($\Delta I(q)$) have been
imported).

Two output files are produced, consisting of the fit (a two-column
ASCII file) and a log file (plain text) showing the values of the
parameters at each iteration of the fitting process, and the final
result including parameter uncertainties, reduced $\chi^2$ and
goodness of fit ($R^2$ value). These are described in more detail in
Section \ref{stats}. The final parameters are also displayed on the
control panel.

Users must be aware of the assumptions made in the modeling of the
data, which uses a hard sphere model with a local monodisperse
approximation. Strongly interacting systems, for example systems
with a high degree of periodicity, are outside the scope of this
approximation. The user is responsible for understanding the
applicability of this approximation to their system, and ensuring
that the fitted results are physically meaningful.

\section{Mathematical details}
\label{sec3}

The small angle scattering intensity $I(q)$ is related to the
scattering cross section $\frac{d\sigma}{d\Omega}(q)$ by

\begin{equation}
I(q) = \phi_0 A t (\Delta\Omega) \frac{d\sigma}{d\Omega}(q)
\label{eq1}
\end{equation}

\noindent where $\phi_0$ is the incident flux (number of photons, or
neutrons, per area per second), $A$ is the illuminated area on the
sample, $t$ is the sample thickness and $\Delta\Omega$ is the solid
angle subtended by a pixel in the detector \cite{GlatterKratky1982}.

For SAXS, the scattering cross-section is calculated from the
structure factor and particle/pore size distribution from

\begin{equation}
\frac{d\sigma}{d\Omega}(q) = r_e^2 (\Delta \rho)^2 N \int_0^{\infty}
n(r) \left[f (qr) \right]^2 S(qr) dr
\label{eq2}
\end{equation}

\noindent where $r_e$ is the electron radius, $\Delta\rho$ is the
electron density contrast, $N$ is the number density, $n(r)$ is the
number fraction particle/pore size distribution (normalized so that
the integral over $r$ is unity), $f(qr)$ is the spherical form
factor, and $S(qr)$ is the structure factor. These terms are defined
in the following sections. The final equation for the intensity used
by the program is

\begin{equation}
I(q) = c \int_0^{\infty} n(r) \left[ f(qr) \right]^2 S(qr) dr
\label{eq3}
\end{equation}

\noindent where the scale factor $c$ is a fitted parameter,
equivalent to

\begin{equation}
c = \phi_0 A t ( \Delta \Omega) r_e^2 (\Delta \rho)^2 N
\label{eq4}
\end{equation}

\noindent For SANS an expression similar to Eq. \ref{eq4} holds.

The data are modeled using a hard-sphere model with local
monodisperse approximation \cite{Pedersen1994}. This model assumes
that the particles are spherical and locally monodisperse in size.
In other words, the particle positions are correlated with their
size. This is a good approximation for systems with large
polydispersity and the approximation provides meaningful results
providing the particles are for the most part not inter-connected
(e.g., the particle concentration is not too high) and are not
spatially periodic. For porous systems (with not too high pore
concentration), the `particle' radius is equivalent to the pore
size.

\subsection{Particle/pore size distribution, $n(r)$}
\label{sec3-1}

The user has the choice of three pore/particle size distributions,
which use the same fitting parameters $r_0$ (`pore size' radius) and
$\sigma$ (`dispersion'). If two size distributions are selected, the
distribution function is expanded to have the form:

\begin{displaymath}
n(r) = (1-\lambda) n_1 (r, r_{0,1}, \sigma_1) + \lambda n_2 (r,
r_{0, 2}, \sigma_2)
\end{displaymath}

\noindent where $\lambda$ is the number fraction of the second
distribution and $r_{0,j}$ and $\sigma_j$ are the $r_0$ and $\sigma$
parameters for the $j^{th}$ distribution. For example, a 50:50
mixture by number fraction would have $\lambda = 0.5$. To model a
situation involving a mixture with differing contrasts, $\lambda$
would be weighted by the different contrast values. The user has a
choice of three distributions, as follows.

\subsubsection{Log normal distribution}

\begin{displaymath}
n(r) = \exp \left( - \frac{1}{2} \frac{\left[ \ln \left(
\frac{r}{r_0}\right)\right]^2}{\sigma^2}\right) \cdot \frac{1}{r
\sigma \sqrt{2 \pi}}
\end{displaymath}

This has a maximum at $r = r_0 \exp \left( - \sigma^2\right)$, a
mean of $r_0 \exp \left(\frac{-\sigma^2}{2}\right)$, and variance
$r_0^2 \left[ \exp \left(2\sigma^2\right) - \exp \left( \sigma^2
\right)\right]$.

\subsubsection{Schulz distribution}

\begin{displaymath}
n(r) = Z^Z X^{Z-1} \frac{\exp(-ZX)}{r \Gamma (Z)}
\end{displaymath}

\noindent where $Z = \frac{1}{\sigma^2}$, $X = \frac{r}{r_0}$, and
$\Gamma(Z)$ is the Gamma function, defined $\forall x \in
\mathbf{R}$ by $\Gamma(x+1) = x \Gamma(x)$, and $\Gamma(1) = 1$
\cite{Lau2004}. The Schulz distribution is frequently used in SANS
analysis. It is physically reasonable in that it is skewed towards
large sizes and has a shape close to a log-normal distribution. As
$Z \rightarrow \infty$ it approaches a Gaussian distribution
\cite{Bartlett1992}.

\subsubsection{Gaussian distribution}

\begin{displaymath}
n(r) = \frac{1}{\sigma \sqrt{2 \pi}} \exp \left[ - \frac{1}{2}
\left( \frac{r-r_0}{\sigma}\right)^2\right]
\end{displaymath}

The Gaussian distribution is symmetric about the mean, $r_0$, and
has variance $(r_0\sigma)^2$. In practise it is only useful for
systems with low polydispersity (small $\sigma$).

\subsection{Spherical form factor, $f(qr)$}
\label{sec3-2}

The spherical form factor has the following form \cite{Pedersen1994,
Kinning1984}:

\begin{displaymath}
f(qr) = r \pi r^3 \left[ \frac{\sin(qr) - \cos(qr) qr}{(qr)^3}
\right]
\end{displaymath}

\subsection{Structure factor, $S(qr)$}

The structure factor follows the local monodisperse approximation
(LMA) for hard spheres \cite{Pedersen1994, Huang2002}, given by

\begin{displaymath}
S(q R_{HS}) = [1 + 24 \eta G(q R_{HS}) / (q R_{HS})]^{-1}
\end{displaymath}

\noindent where $\eta$ is the dimensionless parameter eta (sometimes
referred to as the hard sphere volume fraction, having a value
between 0 and 1), $R_{HS}$ is the hard sphere pore/particle radius,
defined as $R_{HS} = Cr$, where $C$ relates the hard-sphere radius
to the physical particle radius \cite{Pedersen1994}, and
$G(qR_{HS})$ has the form:

\begin{multline}
G(A) = \alpha (\sin A - A \cos A) / A^2 + \\
\beta [2 A \sin A + (2-A^2) \cos A -2] / A^3 + \\
\gamma (- A^4 \cos A + 4 [(3 A^2 - 6) \cos A + \\
(A^3 - 6A) \sin A + 6 ]) / A^5 \notag\\
\end{multline}

\noindent where

\begin{align*}
\alpha & = \frac{(1 _ 2 \eta)^2}{(1 - \eta)^4} \\
 \beta & = -6 \eta \frac{(1 + \eta / 2)^2}{(1 - \eta)^4} \\
\gamma & = \eta \alpha / 2\\
\end{align*}

When a second size distribution is included in the fit, it has the
same $C$-parameter and $\eta$ values as the first distribution. To
set the structure factor to unity, one simply sets $\eta = 0$. This
is appropriate for dilute systems.

\subsection{Fitting routine details}

The program uses a least-squares fitting routine which follows the
Levenberg-Marquardt non-linear regression method to minimize the
reduced $\chi^2$. The integrals are calculated using the Romberg
integration method with $2^{10}$ intervals. Since the integral
\ref{eq1} must have finite bounds on $r$, these are chosen based on
the range of the distribution function $n(r)$, such that $n(r) < 1
\times 10^{-15}$. The bounds are calculated numerically as follows:

For the log-normal and Schulz distributions, the lower bound is $r_0
\exp (-8\sigma)$ and the upper bound $r_0 \exp(8 \sigma)$.

For the Gaussian distribution, the lower bound is the maximum of
zero or $r_0 - 8 \sigma$, and the upper bound is $r_0 + 8 \sigma$.

\subsection{Statistical analysis}
\label{stats}

The reduced $\chi^2$ and $R^2$ (goodness of fit) from the non-linear
regression are reported at the end of the fitting procedure. These
are common statistical measures and defined as follows:

\begin{displaymath}
\mathrm{Reduced} \phantom{a} \chi^2 = \frac{1}{n-p} \sum_i w_i
\left(y_i - F(x_i)\right)^2
\end{displaymath}

\noindent where $n$ is the number of data points, $p$ is the number
of free parameters, $w_i$ are the weightings, $y_i$ is the input
data $I(q)$ and $F(x)$ is the calculated $I(q)$.

\begin{displaymath}
R^2 = 1 - \frac{\sum_i \left( y_i - F(x_i)\right)^2}{\sum_i \left(
y_i - \overline{F(x)}\right)^2}
\end{displaymath}

\noindent where $y_i$ and $F(x_i)$ are defined above, and
$\overline{F(x)}$ is the average of the $F(x)$ values (a constant).

The parameter uncertainties are obtained by calculating the
covariance matrix $C_{ij}$, from

\begin{displaymath}
C = (J^T WJ) ^{-1}
\end{displaymath}

\noindent where $J$ is the Jacobian matrix of partial derivatives
and $W$ is a diagonal matrix where $W_{ii}$ is the weighting on the
$i^{th}$ data point \cite{Toby2004}. Finally the reported parameter
uncertainties are twice the square root of the diagonals of $C,$
i.e. $\delta P_i = 2 \sqrt{C_{ii}}$. This is two standard
deviations, which for a Gaussian distribution of errors represents a
95\% confidence interval.

\section{Examples}

Figure 2 shows examples of data and the fitted result for nanoporous
methyl silsesquioxane films \cite{Huang2002}, formed by spin-coating
a solution of the silsesquioxane along with a sacrificial polymer
(`porogen'), and then annealing to remove the polymer and leave
behind a nanoporous network. As the proportion of porogen is
increased, the pores are observed to increase in size
\cite{Huang2002}. Data are shown for films with porogen loadings of
5 to 25 \% with the background (from methyl silsesquioxane)
subtracted. A single log-normal size distribution was fitted to
each, the $C$-parameter was fixed at 1.1, $\eta$ was fixed to the
porosity (as determined from the porogen loading), and no background
function was used. The results obtained are given in Table
\ref{tab1} and show an increase in the pore size with increased
porogen loadings, in good agreement with electron microscopy and
previous results \cite{Huang2002}.

\begin{figure}
\includegraphics*[width=80mm]{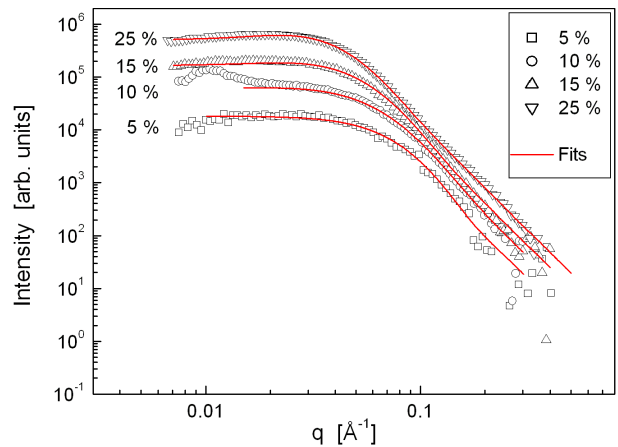}
\caption{\label{fig2} Examples of data from nanoporous
silsesquioxane films, with different porogen loadings. Open symbols:
raw data. Lines: fitted curves using SAXSFit.}
\end{figure}

\begin{table*}
\begin{tabular}{|c|c|c|c|c|}
\hline
Porogen loading&5\%&10\%&15\%&25\%\\
\hline Pore size radius ({\AA})&18.77 $\pm$ 0.14 & 16.94 $\pm$ 0.06
& 22.42 $\pm$ 0.05 & 31.67 $\pm$ 0.06\\
Dispersion & 0.305 $\pm$ 0.004 & 0.382 $\pm$ 0.002 & 0.367 $\pm$
0.001 & 0.370 $\pm$ 0.001 \\
Reduced $\chi^2$ & 1.105 & 2.117 & 2.913 & 2.023 \\
$R^2$ (degree of fit) & 0.9771 &0.9858 &0.9952 &0.9976\\
\hline
\end{tabular}
\caption{\label{tab1} Fitted parameters for nanoporous silsequioxane
film samples shown in Figure \ref{fig2}. The pore radii differ from
that reported by Huang et al. \cite{Huang2002} due to a slightly
different form used for the log-normal distribution. When plotted as
a function of radius the distributions needed to fit the data are
identical.}
\end{table*}

\begin{figure}
\includegraphics*[width=80mm]{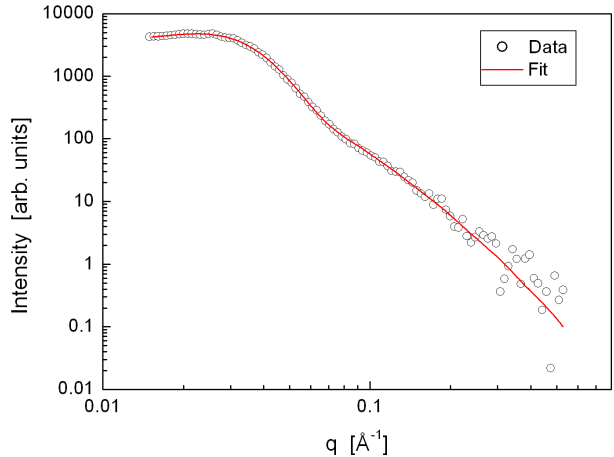}
\caption{\label{fig3} Example of data from nanoporous glass, showing
two pore size distributions. Open symbols: raw data. Line: fitted
curve using SAXSFit.}
\end{figure}

Figure \ref{fig3} shows data from a nanoporous glass using a three
arm star shaped polymer as the porogen \cite{Hedrick2002}, which was
found to exhibit two pore size distributions.

The parameters for the fit are as follows:

First distribution: $r_0 = 50.5 \pm 0.5, \sigma = 0.287 \pm 0.008$.

Second distribution: $r_0 = 15 \pm 1, \sigma = 0.23 \pm 0.05$.

The number fraction of the second distribution was 88 $\pm$ 2 \%,
which equates to a volume of 16 $\pm$ 5 \%. The parameter $\eta$ was
the same for both distributions ($\eta = 0.18 \pm 0.01$) and the
$C$-parameter was fixed at 1.1 for both distributions. The $R^2$
value was 0.9951 and reduced $\chi^2$ 3.33.

\section{Software availability and system requirements}

SAXSFit is provided as a Windows executable (tested on Windows 98,
2000 and XP), or Java .jar executable (tested on Linux Ubuntu and
Mac OS X10.4). The SAXSFit programs and user manual are available
from http://www.irl.cri.nz/SAXSfiles.aspx .

\section{Summary}

SAXSFit is a useful program for fitting small angle x-ray and
neutron scattering data, using a hard sphere model with local
monodisperse approximation. SAXSFit provides an easy-to-use analysis
package for novices and experts. It is stand-alone software and can
be used in a variety of software environments.

\begin{acknowledgments}
Funding was provided in part by the New Zealand Foundation for
Research, Science and Technology under contract CO8X0409. Portions
of this research were carried out at the Stanford Synchrotron
Radiation Laboratory, a national user facility operated by Stanford
University on behalf of the U.S. Department of Energy, Office of
Basic Energy Sciences. The authors also wish to thank Benjamin
Gilbert, Shirlaine Koh, and Eleanor Schofield for testing and
helpful suggestions for improvement, and Peter Ingham for assistance
with the coding.
\end{acknowledgments}

\end{document}